\documentclass[aps,pra,twocolumn,superscriptaddress,showpacs,amsmath,amssymb]{revtex4}

\usepackage{graphicx,graphics,epsfig}
\usepackage{dcolumn}
\usepackage{epstopdf}
\usepackage{bm,amsmath,amssymb}
\begin{document}

\newcommand{\ve}{\varepsilon}
\newcommand{\s}{\sigma}
\newcommand{\de}{q}
\newcommand{\D}{q}
\newcommand{\g}{\gamma}
\newcommand{\al}{\alpha}
\newcommand{\be}{\beta}
\newcommand{\ka}{\kappa}
\newcommand{\om}{\omega}
\newcommand{\vp}{\varphi}
\makeatletter
\newcommand{\rmnum}[1]{\expandafter\@slowromancap\romannumeral #1@}
\makeatother

\title{Nonreciprocal Anderson Localization in Magneto-Optical Random Structures}

\author{Konstantin Y. Bliokh}
\affiliation{Applied Optics Group, School of Physics, National University of Ireland, Galway, Galway, Ireland}
\affiliation{A. Usikov Institute of Radiophysics and Electronics, 12 Ak. Proskury St., Kharkov 61085, Ukraine}

\author{Sergey A. Gredeskul}

\affiliation{Department of Physics, Ben Gurion University of the Negev, Beer Sheva 84105, Israel}

\author{Puvanesvari Rajan}

\affiliation{Nonlinear Physics Centre, Research School of Physics and Engineering,
The Australian National University, Canberra, ACT 0200, Australia}

\author{Ilya V. Shadrivov}

\affiliation{Nonlinear Physics Centre, Research School of Physics and Engineering,
The Australian National University, Canberra, ACT 0200, Australia}

\author{Yuri S. Kivshar}

\affiliation{Nonlinear Physics Centre, Research School of Physics and Engineering,
The Australian National University, Canberra, ACT 0200, Australia}

\begin{abstract}
We study, both analytically and numerically, disorder-induced localization of light in random layered structures with magnetooptical materials. The Anderson localization in such structures demonstrates nonreciprocal features in the averaged localization length and individual transmission resonances. We employ short-wavelength approximation where the localization effects are strong, and consider both the Faraday and Voigt magnetooptical geometries. In the Faraday geometry, the transmission is strongly nonreciprocal for the circularly polarized waves, whereas in the Voigt geometry, the nonreciprocity is much weaker, and it may appear only for the individual transmission resonances of the TM-polarized waves.
\end{abstract}


\maketitle

\section{Introduction}
\label{sec:intro}

Anderson localization is a well known phenomenon associated with wave interference arising from multiple scattering by defects. Being originally suggested for the suppression of classical diffusive motion of electrons due to disorder introduced into a periodic structure~\cite{Anderson}, the Anderson localization is currently associated with many physical systems of different origin and nature. As a matter of fact, the Anderson localization is universal to all wave systems in disordered potentials and media (see, e.g., Refs.~\cite{PingSheng,FreilikherGredeskul,BaluniWillemsen,BEC} to cite a few). In particular, the Anderson localization occurs upon propagation and scattering of light in random media, which are microscopically transparent but appear opaque because of the localization effect~\cite{PingSheng,FreilikherGredeskul,john,BerryKlein}. Many experimental studies of the localization properties of light include the demonstration of exponential decay of transmittance with the sample length~\cite{wiersma-genack} and the study of transverse localization 
in one- and two-dimensional disordered photonic lattices~\cite{devries-segev-silberberg}.

Here we study, both analytically and numerically, the Anderson localization of light propagating through random \textit{magnetoactive}
layered structures. We demonstrate that an interplay between strong localization and magnetooptical effects produces a number of \textit{nonreciprocity} features in the transmission characteristics.

Magnetooptical effects and nonreciprocity are widely exploited in modern optics and applied physics~\cite{Zvezdin,Potton}. In particular, magnetoactive periodic structures are currently attracting growing attention \cite{Lubchenskii,Inoue}. The main phenomena of interest are the enhanced Faraday effect on resonances~\cite{InoueAraiFujii} and one-way propagation (nonreciprocal transmission) \cite{FigotinVitebsky,YuWangFan,Khanikaev,KhanikaevSteel} employed for the concept of optical insulators. The resonant Faraday effect has also been shown in connection with the localization of light in random layered structures~\cite{InoueFujii}. Although the destructive role of the Faraday effect on the weak localization effect in three-dimensional random scattering media was examined previously~\cite{ErbacherLenkeMaret}, there is no analysis of the strong Anderson localization in random magnetoactive media.

In this paper, we examine the transmission properties of one-dimensional random layered structures with magnetooptical materials.
We employ short-wavelength approximation, where the localization is strong, and consider both Faraday and Voigt geometries. In  the Faraday geometry, magnetooptical correction to the \textit{localization length} results to a significant \textit{broadband nonreciprocity} and polarization selectivity in the typical, \textit{exponentially small transmission}. In the Voigt geometry, averaging over random phases suppresses the magnetooptical effect, in contrast to the case of periodic structures where it can be quite pronounced~\cite{FigotinVitebsky,KhanikaevSteel}. At the same time, in both the geometries we reveal the nonreciprocal frequency shifts of narrow \textit{transmission resonances}, corresponding to the excited \textit{localized states} inside the structure \cite{Frish,BliokhBliokhFreilikher}. This offers efficient \textit{unidirectional propagation} at the given resonant frequency.

The paper is organized as follows. In Section~II we introduce general formalism for the analysis of the wave propagation and localization in random layered structures, and also discuss generic aspects of nonreciprocity. Section~III is devoted to the explicit calculations of the averaged localization lengths in the Faraday and Voigt magnetooptical structures. Transmission resonances in these structures are discussed in Section~IV.
Finally, Section~V concludes the paper.

\section{General Formalism}
\label{sec:formulation}
\subsection{Basic equations}
\label{subsec:basic}

We consider transmission of a polarized electromagnetic wave incident on a random stack of the length $L$ which consists of ${\mathcal N}$ dissipationless magnetoactive layers possessing random widths and different optical parameters. The widths of the layers are independent random values with mean value $\bar{w}=L/{\mathcal N}$ and variance $d$. Figure~1 displays a scheme of the system with two alternating types of the layers. Here the stack is formed by a sequence of ${\mathcal N}=2N$ layers of two different types labeled by indices `$a$' and `$b$' and the whole structure is surrounded by vacuum labeled by index `$0$'. The $z$-axis is directed across the layers.

In the simplest case of one type of propagating waves, the waves at each point can be described by two amplitudes
$h^{\upsilon}$ corresponding to the propagation in positive $\upsilon=1$ and negative  $\upsilon=-1$ directions with respect to $z$-axis. Transmission of these waves through the structure is described by the total transfer matrix ${\hat T}$, which expresses the input amplitudes via their output value:
%
\begin{equation}
\label{total trans matr}
    {\vec{h}}(0)={\hat T}\, \vec{h}(L)~,
    \ \ \
    \vec{h}\equiv\left(
    \begin{array}{c}
              h^+ \\
              h^-
    \end{array}
            \right).
\end{equation}
The amplitudes are assumed to be normalized so that the intensity $|h|^2$ gives the wave energy. Then the transfer matrix is unimodular,
$\det {\hat T} =1$, which ensures the energy flux conservation.

The transmission coefficient $T$ for the wave $h^+$ incident on the system from the left is simply related to the first diagonal element of the transfer matrix:
%
\begin{equation}
\label{total-transm-coeff}
    T=\frac{1}{\left({\hat T}\right)_{11}}~.
\end{equation}
The corresponding transmittance of the structure is
%
\begin{equation}
\label{total-transm}
    {\mathcal T}=|T|^2~.
\end{equation}

\begin{figure}[tbh]
\centering
\includegraphics[width=0.9\columnwidth]{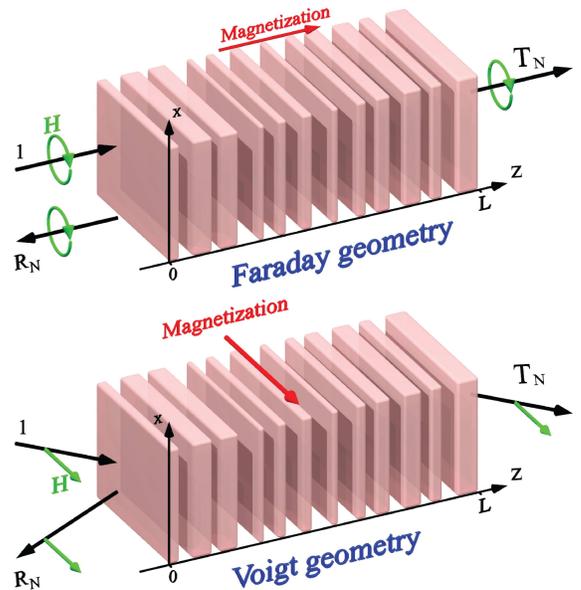}
\caption{\label{fig:stack}(Color online.) Schematic picture of the wave transmission and reflection from a random-layered structure consisting of two types of alternating layers `$a$' (here -- a magnetoactive material) and `$b$' (here -- air) with random widths. Magnetization of the medium, wave polarizations and directions of propagation are shown for the Faraday and Voigt geometries.}
\end{figure}

Due to Anderson localization, the absolute value of the transmission coefficient exponentially decreases with the stack length for a typical random realization. Such a decrease is described by the dimensionless transmission decrement \cite{BaluniWillemsen,BerryKlein}
%
\begin{equation}
\label{transm decrement}
    \kappa=-
    \frac{\left<\ln {\mathcal T}\right>}{2N}~,
\end{equation}
where $\left< ... \right>$ stands for the ensemble averaging. In the case of $N\kappa\gg 1$, it coincides with the inverse dimensionless localization length $\ell_{\rm loc}$:
%
\begin{equation}
\label{loclength}
    \kappa\approx\frac{1}{\ell_{\rm loc}}=-\lim_{N\to\infty}\frac{\ln {\mathcal T}}{2N}=-\lim_{N\to\infty}\frac{\left<\ln {\mathcal T}\right>}{2N}.
\end{equation}
The real dimensional localization length is obtained by multiplication of $\ell_{\rm loc}$ by the averaged length of one period, $2\bar{w}$.

We are interested in two transmission characteristics: transmittance on a realization, ${\mathcal T}$, Eq. (\ref{total-transm}), and transmission decrement $\kappa$, Eq. (\ref{transm decrement}). Following Ref.~\cite{BerryKlein}, we represent our system as a sequence of uniform layers of media `$a$' or `$b$' and interfaces between them (Fig.~\ref{fig:stack}). Wave propagation inside the layers is described by diagonal transfer matrices:
%
\begin{equation}
\label{space}
    {\hat S}_k=\left(
    \begin{array}{cc}
      {\textrm e}^{-i\varphi_k} & 0 \\
      0 & {\textrm e}^{i\varphi_k}
    \end{array}
    \right),
\end{equation}
where $\varphi_k$ is the phase accumulated upon the wave propagating from left to right through the $k$-th layer. Here the layers with odd numbers $k=2j-1, \ \ j=1,2,...,N$ are filled with medium `$a$', whereas those with even numbers  $k=2j, \ \ j=1,2,...,N$ are filled with the medium `$b$'.

The interfaces are described by unimodular transfer matrices ${\hat F}^{0a},{\hat F}^{ab},{\hat F}^{ba},{\hat F}^{a0}$ corresponding, respectively, to transitions from vacuum to the medium `$a$', from the medium `$a$' to the medium `$b$', from the medium `$b$' to the medium `$a$', and from the medium `$a$' to vacuum. Thus, the total transfer matrix (\ref{total trans matr}) of the structure is
%
\begin{eqnarray}
\label{total transf matr}
    \hat{T}={\hat F}^{0a}{\hat S}_1{\hat F}_{1}{\hat S}_2{\hat F}_{2}~...~
    {\hat S}_{2N-1}{\hat F}_{2N-1}{\hat S}_{2N}{\hat F}_{2N}{\hat F}^{a0},\nonumber\\
    {\hat F}_{2j-1}\equiv {\hat F}^{ab}, \ \ {\hat F}_{2j}\equiv {\hat F}^{ba}, \ j=1,2,...,N.
\end{eqnarray}
Here, in the two last multipliers, we used the group property of the interface transfer matrices: ${\hat F}^{b0}={\hat F}^{ba}{\hat F}^{a0}$.

Our numeric calculations of transmittance (\ref{total-transm}) and transmission decrement (\ref{transm decrement}) are based on the exact expression (\ref{total transf matr}). At the same time, analytical expression for the transmission decrement can be readily obtained in the short-wavelength approximations. 

First, in the localized regime, the main contribution to the localization length is provided by $2N$ transfer matrices of the layers and we can neglect the external interface transfer matrices ${\hat F}^{0a}$, ${\hat F}^{a0}$ just replacing the exact matrix $\hat{T}$ by the truncated matrix $\hat{T}^{\prime}$:
%
\begin{equation}
\label{truncated}
    \hat{T}^{\prime}={\hat S}_1{\hat F}_{1}{\hat S}_2{\hat F}_{2}~...~
    {\hat S}_{2N-1}{\hat F}_{2N-1}{\hat S}_{2N}{\hat F}_{2N}.
\end{equation}
Second, if the wavelength within the $k$-th layer is much shorter than the variance of the layer thikness \cite{BerryKlein}:
\begin{equation}
\label{short-wave}
    \frac{\lambda_k}{2} \ll d~,
\end{equation}
then the phases $\varphi_k$ modulo $2\pi$ in the propagation matrices (\ref{space}) are nearly uniformly distributed in the range $(0,2\pi)$. In this approximation, the transmittance for the transfer matrix (\ref{truncated}) averaged over the phases $\varphi_k$ is reduced to the product of the transmittances of separate layers \cite{BaluniWillemsen} and, furthermore, to the product of transmittances of the interfaces only \cite{BerryKlein}:
%
\begin{equation}
\label{factorization}
    \ln \left({\mathcal T}\right)\approx
    \sum_{j=1}^{2N} \ln \tau_j~,~~\tau_j=1/|({\hat F}_j)_{11}|^2.
\end{equation}
Substitution of Eq.~(\ref{factorization}) into Eq.~(\ref{loclength}) yields the following simple expression for the transmission decrement and corresponding localization length:
%
\begin{equation}
\label{transm decrement-1}
    \kappa=\ell_{\rm loc}^{-1}=
    {\ln\left|\left({\hat F}^{ab}\right)_{11}\left({\hat F}^{ba}\right)_{11}\right|}.
\end{equation}

This is the main result for calculation of the localization length in the short-wavelength approximation. It can be easily extended to any number of alternating layers. For instance, considering a random structure consisting of three types of alternating layers, `$a$', `$b$', and `$c$', with ${\mathcal N}=3N$, one has
\begin{equation}
\label{3-layer}
    \kappa={\ln\left|\left({\hat F}^{ab}\right)_{11}\left({\hat F}^{bc}\right)_{11}\left({\hat F}^{ca}\right)_{11}\right|}.
\end{equation}
%
\subsection{Nonreciprocal transmission}
\label{subsec:non-reciprocity}
Transmission through a one-dimensional dissipationless linear medium is always reciprocal if there is only one (but propagating in two directions) mode in the system. Indeed, while the forward transmission of the $h^+$ wave incident from the left on the medium is described by the $2\times 2$ transfer matrix ${\hat T}$, the backward transmission of the reciprocal $h^-$ wave incident from the right is characterized by the inverse transfer matrix ${\hat T}^{-1}$. Using the ${\rm SU}(1,1)$ group properties of the transfer matrix, one can easily show that the corresponding transmission coeffisient (\ref{total-transm-coeff}) of the backward wave is $T$, i.e., exactly coincides with that of the forward wave \cite{BaluniWillemsen,BerryKlein}.

If the system possesses two or more uncoupled modes labeled by index $\sigma$, the situation can be more complicated. In this case, the waves are marked by the propagation direction and mode indices: $h^{\upsilon,\sigma}$. Still, the forward and backward propagation of each mode $\sigma$ through the system ($h^{+,\sigma}$ and $h^{-,\sigma}$ incident waves) are described by the $2\times 2$ transfer matrices ${\hat T}^{\sigma}$ and $({\hat T}^{\sigma})^{-1}$ characterized by the same transmittance $\mathcal{T}^\sigma$. However, the wave reciprocal to $h^{+,\sigma}$ is determined by the \textit{time-reversal operation} which changes $\upsilon \mapsto -\upsilon$ (because of the $\bf{k}\mapsto -\bf{k}$ transformation) but can also affect $\sigma$ \cite{Potton}. In particular, if the time reversal operation changes the sign of the mode index: $\sigma \mapsto -\sigma$, then the \textit{reciprocal} wave will be $h^{-,-\sigma}$ rather than the \textit{backward} wave of the same mode, $h^{-,\sigma}$. Accordingly, the transmittances of the mutually reciprocal waves through the system, $\mathcal{T}^\sigma$ and $\mathcal{T}^{-\sigma}$, can be different. This signals \textit{nonreciprocity} in the system.

Note that noreciprocity in the system under consideration originates from the difference between the modes $\sigma$ and $-\sigma$, and does not depend explicitly on the direction of incidence $\upsilon$. Therefore, in practice, it is sufficient to compare only \textit{forward} transmissions of the modes $\pm\sigma$, described by the transfer matrices ${\hat T}^{\pm\sigma}$ and transmittances $\mathcal{T}^{\pm\sigma}$.

As we show below, propagation of light in magnetoactive layered media offers nonreciprocal transmission which can be explained within the above formalism. There are two main geometries typical for magneto-optical problems \cite{Zvezdin}: (i) the Faraday geometry, where the magnetization is collinear with the direction of propagation of the wave, and (ii) the Voigt (or Cotton-Mouton) geometry, where the magnetization is orthogonal to the direction of propagation of the wave (see Fig.~1). In the next Sections we study the averaged transmission decrement (Sec. III) and individual transmission resonances (Sec. IV) in both geometries.
\section{Localization decrements}
\label{sec:Localization}
\subsection{Faraday geometry}
\label{subsec:FaradayLocalization}
In the Faraday geometry both magnetization and the wave vector are directed across the layers, i.e., along the $z$-axis (see Fig.~1). We assume that the magneto-optical effects are described exclusively by the dielectric tensor, while the magnetic tensor is equal to one. In this case, the electric induction in the medium reads \cite{Zvezdin}
%
\begin{equation}
\label{induction}
    {\bf D} =\varepsilon{\bf E}+i{\bf g}\times{\bf E}\equiv\hat{\varepsilon}\,{\bf E}~, \ \ {\bf g}=\left(0,0, Q\right),
\end{equation}
where ${\bf E}$ is the electric field of the wave, $\varepsilon$ is the isotropic dielectric constant in the absence of magnetization, and $Q\equiv\varepsilon q$ is the magneto-optical constant proportional to the magnetization of the medium. In what follows, we assume that $|q|\ll 1$ and will be interested in the effects linear in $q$.

Thus, the dielectric tensor in the Faraday geometry has the form:
%
\begin{equation}
\label{vareps}
    \hat{\varepsilon}=\varepsilon
  \left(%
    \begin{array}{ccc}
    1 & -iq & 0 \\
    iq &  1 & 0 \\
    0 & 0 & 1\\
    \end{array}%
    \right).
\end{equation}
Solving stationary Maxwell equations for the magnetic field ${\bf H}=(H_{x},H_{y},0)$ in a homogeneous medium,
%
\begin{equation}
\label{Maxwell-1}
    -{\bf k}\times\left[\hat{\varepsilon}^{-1}
    \left({\bf k}\times \textbf{H}\right)\right]
    =k_{0}^{2}\,{\bf H}~,
\end{equation}
$k_{0}=\omega/c$, we find that the eigenmodes of the problem are circularly polarized waves:
%
\begin{equation}
\label{magnetic}
    {\bf H}^{\upsilon,\sigma}=\frac{H^{\upsilon,\sigma}}{\sqrt{2}}
    \left(%
    \begin{array}{c}
    1\\
    i\sigma\\
    0\\
    \end{array}%
    \right){\textrm e}^{ i(\upsilon kz-\omega t)}, \ \ \ \upsilon,\sigma=\pm 1.
\end{equation}
The corresponding wave electric field ${\bf E}$ is
%
\begin{equation}
\label{electric}
    {\bf E}^{\upsilon,\sigma}= i\upsilon\sigma \frac{k_0}{k}\bf{H}^{\upsilon,\sigma}.
\end{equation}
In these equations, $H^{\upsilon,\sigma}$ are the wave amplitudes, whereas
%
\begin{equation}
\label{wavenumbers}
    k=nk_{0}\sqrt{1+\sigma q}, \ \ \
    n=\sqrt{\varepsilon},
\end{equation}
is the propagation constant affected by the magnetization parameter $q$ and depending on $\sigma$. In the linear approximation in $q$, $k\simeq nk_{0}(1+\sigma q/2)$.

Parameter $\sigma$ is the mode index which determines the direction of rotation of the wave field. In this manner, the product $\upsilon\sigma$ represent the helicity
%
\begin{equation}
\label{helicity}
	\chi=\upsilon\sigma,
\end{equation}
which distinguishes the right-handed ($\chi=+1$) and left-handed ($\chi=-1$) circular polarizations defined with respect to the direction of propagation of the wave. Note that the time reversal operation keeps helicity unchanged, whereas $\sigma$ changes its sign \cite{Potton}. Thus, the reciprocal wave is given by ${\bf H}^{-\upsilon,-\sigma}$, precisely as described in Section IIB.

Consider the wave transformation at the interface between the media `$a$' and `$b$'. The helicity of the wave flips upon the reflection and remains unchanged upon transmission. As a result, parameter $\sigma$ remains unchanged, so that there is no coupling between the modes with $\sigma=+1$ and  $\sigma=-1$ (see Fig.~1), and these modes can be studied independently. From now on, for the sake of simplicity, we omit $\sigma$ in superscripts and write explicitly the values of the direction parameter $\upsilon=\pm 1$. In this manner, the boundary conditions for the wave electric and magnetic fields at the `$a$'-`$b$' interface read
%
\begin{equation}
\label{boundary-1}
    {\bf H}_{a}^{+}+{\bf H}_{a}^{-}={\bf H}_{b}^{+}+{\bf H}_{b}^{-},
     \ \ \ \
    {\bf E}_{a}^{+}+{\bf E}_{a}^{-}={\bf E}_{b}^{+}+{\bf E}_{b}^{-}.
\end{equation}
Substituting Eqs.~(\ref{magnetic}) and (\ref{electric}) into Eqs.~(\ref{boundary-1}), we obtain that the wave amplitudes in the two media, $\vec{H}_{a,b}\equiv\left( H_{a,b}^+,H_{a,b}^- \right)^T$, are related through the unnormalized interface transfer matrix:
%
\begin{eqnarray}
\label{transfer-1-1}
    \vec{H}_{a}=
    {\hat {\tilde F}^{ab}}
    \vec{H}_{b}~,~~{\hat {\tilde F}^{ab}}=\frac{1}{2k_b}
    \left(%
    \begin{array}{cc}
    \displaystyle{k_b+k_a} & \displaystyle{k_b-k_a}\\
    \displaystyle{k_b-k_a} & \displaystyle{k_b+k_a}\\
    \end{array}%
    \right).
   \end{eqnarray}
Here $k_{a,b}$ are the wave numbers (\ref{wavenumbers}) in the corresponding media.

The determinant of the matrix (\ref{transfer-1-1}), $\det{\hat {\tilde F}^{ab}}=k_a/k_b$, determines the choice of the normalized amplitudes
%
\begin{equation}
\label{normalized field}
    \vec{h}_1=\frac{k_0}{k_a}\vec{H}_1, \ \ \
    \vec{h}_2=\frac{k_0}{k_b}\vec{H}_2,
\end{equation}
and the normalized interface transfer matrix is
%
\begin{equation}
\label{normalized matrix-Faraday}
    {\hat F}^{ab}=\frac{1}{2\sqrt{k_{a} k_{b}}}
    \left(%
    \begin{array}{cc}
    \displaystyle{k_{b}+k_{a}} & \displaystyle{k_{b}-k_{a}}\\
    \displaystyle{k_{b}-k_{a}} & \displaystyle{k_{b}+k_{a}}\\
    \end{array}%
    \right).
\end{equation}
%

Considering now a random multi-layer structure and calculating the localization decrement from Eq. (\ref{transm decrement-1}) with Eqs.~(\ref{wavenumbers}) and (\ref{normalized matrix-Faraday}), we obtain in the linear approximation in $q$:
%
\begin{eqnarray}
\label{transm decrement-2}
    &&\kappa=2\ln\frac{k_{a}+ k_{b}}{2\sqrt{k_a k_b}}\simeq \kappa^{(0)}+\kappa^{(1)},\nonumber\\
    &&\kappa^{(0)}\!=\!\ln\frac{(n_a+n_b)^2}{4n_a n_b},\nonumber\\  &&\kappa^{(1)}\!=\!\frac{\sigma}{2}
    (q_a-q_b)\frac{n_a-n_b}{n_a+n_b}.
\end{eqnarray}
Thus, the localization decrement acquires the first-order magneto-optical correction $\kappa^{(1)}$ caused by the Faraday effect. This correction depends on $\sigma$, i.e., on the polarization helicity $\chi$ and the propagation direction $\upsilon$ through $\sigma=\chi\upsilon$. For the reciprocal waves with the same $\chi$ and opposite $\upsilon$, $\kappa^{(1)}$ has opposite signs. This signals \textit{nonreciprocal localization} in a Faraday random medium. In practice, the nonreciprocal difference in the transmission decrements (\ref{transm decrement-2}) can be observed by changing sign of either propagation direction $\upsilon$ (with the helicity fixed), or polarization $\chi$, or magnetization $q$.

\begin{figure}[t]
\centering
\includegraphics[width=0.8\columnwidth]{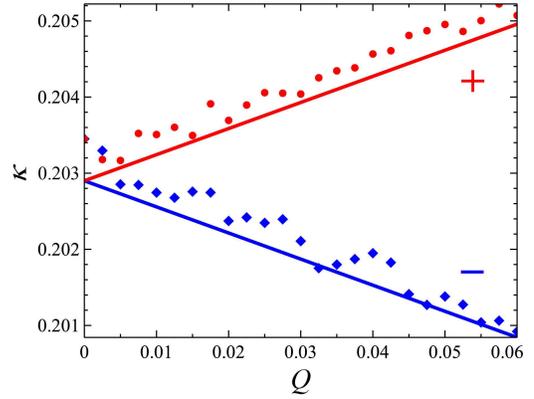}
\caption{\label{fig:Fig2}(Color online.) Localization decrement $\kappa$ vs. magneto-optical parameter $Q$ for opposite modes propagating through a two-component random structure in the Faraday geometry (see details in the text). The modes with $\sigma=\pm 1$ correspond to either opposite circular polarizations or propagation directions. Numerical simulations of exact equations (symbols) and theoretical formula (\ref{transm decrement-2}) (lines).}
\end{figure}

Despite the magneto-optical correction to the localization decrement is small in magnitude, $\kappa^{(1)} \ll \kappa^{(0)}$, it still might result in a significant difference in the typical transmission spectrum. This difference is described by an additional factor of $\propto \exp[-2N\kappa^{(1)}]$ in transmittance, which is exponential with respect to the length of the structure. Hence, small  correction (\ref{transm decrement-2})  brings about significant broadband nonreciprocity or polarization selectivity in the typical small transmission when $N\left|\ka^{(1)}\right|\geq 1$.

Fig.~\ref{fig:Fig2} shows dependence of the localization decrement on the magnetization parameter $Q=\varepsilon q$ calculated numerically and compared to analytical result (\ref{transm decrement-2}). Numerical simulations were performed for the structure containing ${\mathcal N}=2N=90$ alternating layers of air ($\varepsilon = 1, Q=0$), and bismuth iron garnet (BIG), with dielectric constant $\varepsilon = 6.25$ and magneto-optic parameter reaching $Q=0.06$. The thicknesses of layers were randomly distributed in the range  50$\div$150 $\mu$m (i.e., $\bar{w}=100\mu$m, $d=50\mu$m), whereas the excitation wavelength was 632 nm. The averaging was performed over $10^5$ realizations of the random sample. One can see excellent agreement between numerical simulations and analytical results showing linear splitting of the $\sigma=1$ and $\sigma=-1$ localization decrements as a function of the magneto-optic parameter.
\subsection{Voigt geometry}
\label{subsec:Voigt}
In Voigt geometry the magnetization is directed orthogonally to the $z$-axis, say, along the $y$-axis. Then, the dielectric tensor is
%
\begin{equation}
\label{varepsilon-V}
	\hat{\ve}=\ve\left(%
    \begin{array}{ccc}
    1&0&i\de\\
    0&1&0\\
    -i\de&0&1\\
    \end{array}%
    \right).
\end{equation}
The first-order interaction of the wave with the magnetization occurs only upon \textit{oblique} propagation of the wave in the $xz$-plane, i.e., when $k_{x}=\textrm{const} \neq 0$ (see Fig.~1). The eigenmodes in such problems are the TE and TM linearly-polarized modes. Since the TE mode is uncoupled from the magnetization and effectively propagates as in isotropic layered medium, we consider only non-trivial TM polarization. Solving Maxwell equations (\ref{Maxwell-1}) with dielectric tensor (\ref{varepsilon-V}) we obtain that the wave magnetic field of the TM mode is directed along the $y$-axis whereas the electric field has $x$- and $z$-components. For transmission and localization properties of the structure, only components tangential to the layer interfaces are important:
%
\begin{eqnarray}
\label{fields}
    H_y^{\upsilon,\sigma} &=& H^{\upsilon,\sigma} \ \textrm {e}^{i(\sigma x k_{\perp}  + \upsilon z k_{\parallel}-\om t)},\nonumber\\
    E_x^{\upsilon,\sigma} &=& A^{\upsilon,\sigma} H_y^{\upsilon,\sigma}.
\end{eqnarray}
Here parameters $\upsilon=\pm 1$ and $\sigma=\pm 1$ indicate propagation in the positive and negative $z$ and $x$ directions, respectively, $k_{\parallel}=\sqrt{k^2-k_x^2}$, $k_{\perp}=|k_x|$, whereas
%
\begin{equation}
\label{wavenumbers-1}
    A^{\upsilon,\sigma}=\frac{i\sigma q k_{\perp}+ \upsilon k_{\parallel}}{\ve(1-\de^2)k_0},~k=nk_{0}\sqrt{1-\de^{2}}.
\end{equation}
Note that $A^{\upsilon,\sigma}\simeq (\upsilon k_{\parallel} +i\sigma q k_{\perp})/(\varepsilon k_0)$ and $k\simeq nk_0$ in the linear approximation in $q$, so that the magnetization affects \textit{imaginary} parts (i.e., \textit{phases}) of the amplitudes $A^{\upsilon,\sigma}$ and does not affect the propagation constant, cf. Eqs.~(\ref{electric}) and (\ref{wavenumbers}).

In the Voigt geometry, direction of the transverse wave vector component, $\sigma$, serves as the mode index. The mutually reciprocal waves are $H^{\upsilon,\sigma}$ and  $H^{-\upsilon,-\sigma}$ because the time reversal transformation reverts the whole wave vector, $\bf{k}\mapsto -\bf{k}$. Thus, we again deal with a formalism described in Section IIB.

Evidently, the parameter $\sigma$ is not changed upon reflection and transmission through the layers, i.e., the modes with $\sigma=\pm 1$ are uncoupled from each other. Therefore, for the sake of simplicity, we omit the mode index in superscripts, and write explicitly only the values of the direction parameter $\upsilon = \pm 1$. Matching the tangential components of the fields (\ref{fields}) at the interface between media `$a$' and `$b$', Eq.~(\ref{boundary-1}), we obtain the unnormalized transfer matrix relating the wave amplitudes in two media, $\vec{H}_{a}= \hat{{\tilde F}}^{ab}\vec{H}_{b}$ \cite{KhanikaevSteel}:
%
\begin{equation}
\label{transfer-matrix-2}
    \hat{{\tilde F}}^{ab}=\frac{1}{A_{a}^{+}-A_{a}^{-}}\left(
    \begin{array}{ccc}
    {A_{b}^{+}-A_{a}^{-}} & {A_{b}^{-}-A_{a}^{-}}\\
     {A_{a}^{+}-A_{b}^{+}}& {A_{a}^{+}-A_{b}^{-}}\\
    \end{array}
    \right).
\end{equation}
Here $A_{a,b}^{\pm}$ are the amplitudes (\ref{wavenumbers-1}) in the corresponding medium. Noticing that $A^- = A^{+*}$, we calculate the determinant of the matrix (\ref{transfer-matrix-2}), $\det{\hat {\tilde F}^{ab}}={\rm Re}A_{b}^{+}/{\rm Re} A_{a}^{+}$, which determines the normalized field
%
\begin{equation}
\label{normalized fieldVoigt}
    {\vec{h}}_a=\sqrt{2{\rm Re}A_{a}^{+}}\; \vec{H}_a~, \ \ {\vec{h}}_b=\sqrt{2{\rm Re}A_{b}^{+}}\; \vec{H}_b~.
\end{equation}
As a result the normalized transfer matrix takes the form
%
\begin{eqnarray}
\label{normalized matrix-Voigt}
    {\hat F}^{ab}=\frac{1}{\sqrt{
    4{\rm Re}A_{a}^{+}{\rm Re}A_{b}^{+}}}
    \left(%
    \begin{array}{cc}
    \displaystyle{A_{b}^{+}+A_{a}^{+*}} & \displaystyle{A_{a}^{+*}-A_{b}^{+*}}\\
    \displaystyle{A_{a}^{+}-A_{b}^{+}} & \displaystyle{A_{a}^{+}+A_{b}^{+*}}\\
    \end{array}%
    \right).
\end{eqnarray}
%
\begin{figure}[t]
\centering
\includegraphics[width=0.8\columnwidth]{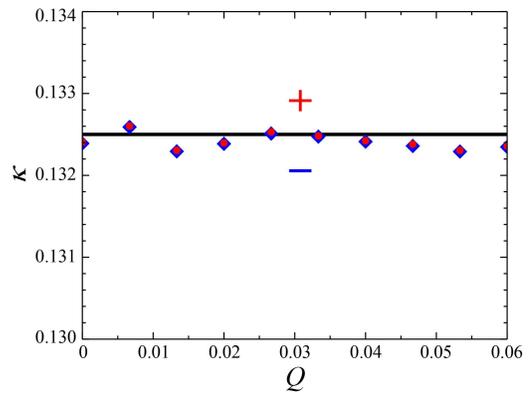}
\caption{\label{fig:Fig3}
(Color online.) Localization decrement $\kappa$ vs. magneto-optical parameter $Q$ for reciprocal waves with $\sigma=\pm 1$ propagating through a three-component magnetooptical random structure in the Voigt geometry (see details in the text). Numerical simulations of exact equations (symbols) and theoretical formula (\ref{transm decrement-4}) (line).}
\end{figure}

In contrast to the Faraday geometry, in the Voigt geometry the linear magneto-optical correction changes only \textit{phases} of the transmission and reflection coefficients, whereas corrections to the interface transmittance start with the terms $\propto q^2$. In short-wave limit, only these transmittances determine the total transmittance, Eq.~(\ref{factorization}). Therefore, a short-wavelength transmission through a random multilayered stack is reciprocal and is not affected by magnetization in the first-order approximation. In particular, substituting Eqs.~(\ref{wavenumbers-1}) and (\ref{normalized matrix-Voigt}) into Eq.~(\ref{transm decrement-1}), we arrive at the localization decrement for a two-component random layered structure (cf. Ref.~\cite{Oblique}):
%
\begin{equation}
\label{transm decrement-3}
    \kappa = \ln {{\left( {\varepsilon _1 k_{2 \parallel}  + \varepsilon _2 k_{1 \parallel} } \right)^2 } \over {4\varepsilon _1 \varepsilon _2 k_{1 \parallel} k_{2 \parallel} }}+O\left(q^2 \right),
\end{equation}
where $k_{\parallel}\simeq \sqrt{n^2 k_{0}^{2}-k_x^2}$. Obviously, it depends on neither propagation nor magnetization directions.

\begin{figure}[t]
\centering
\includegraphics[width=0.83\columnwidth]{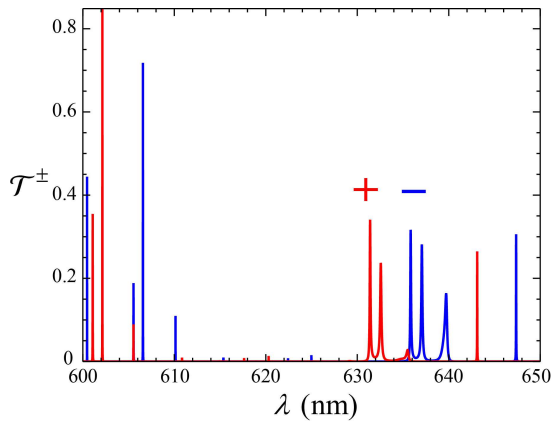}
\caption{\label{fig:faraday_spectrum}
(Color online.) Transmission spectra of a random magneto-optical sample in the Faraday geometry (see details in the text) for waves with $\sigma=\pm 1$. While the averaged localization decrements are only slightly different (Fig.~2), all individual resonances are shifted significantly as compared with their widths, Eq.~(\ref{frequency shift-1}).}
\end{figure}

It is worth remarking that transmission through a \textit{periodic} structure consisting of the two types of alternating layers is also reciprocal in the Voigt geometry for all wavelengths. However, a periodic structure with a cell consisting of \textit{three} different layers (which breaks the mirror reflection symmetry) can demonstrate significant nonreciprocity \cite{YuWangFan,KhanikaevSteel}. At the same time, the short-wavelength localization in a random-layered structure is still reciprocal for any number of components. This is because of independent action of all the interfaces after the phase averaging, Eq.~(\ref{factorization}). In particular, considering a random structure consisting of three types of alternating layers, `$a$', `$b$', and `$c$', we substitute Eqs.~(\ref{wavenumbers-1}) and (\ref{normalized matrix-Voigt}) into Eq.~(\ref{3-layer}) and obtain the localization decrement in the first-order approximation in $q$:
%
\begin{equation}
\label{transm decrement-4}
\kappa \simeq \ln {{\left( {\varepsilon_a k_{\parallel b}  + \varepsilon_b k_{\parallel a} } \right)\left( {\varepsilon_b k_{\parallel c}  + \varepsilon_c k_{\parallel b} } \right)\left( {\varepsilon_c k_{\parallel a}  + \varepsilon_a k_{\parallel c} } \right)} \over {8 \varepsilon_a \varepsilon_b \varepsilon_c k_{\parallel a} k_{\parallel b} k_{\parallel c} }}.
\end{equation}

We verified this result numerically as it is shown in Fig.~\ref{fig:Fig3}. There we calculated forward and backward, $\upsilon=\sigma=\pm 1$, transmission through ${\mathcal N}=3N=60$ random-width layers of three alternating types: air, BIG, and glass ($n=1.5$). The excitation wavelength was 632 nm and the angle of incidence $\theta_0 \equiv \sin^{-1}(k_{\perp}/k_0)= 23^{\circ}$. The layer thicknesses were randomly distributed in the range 50$\div$150 $\mu$m ($\bar{w}=100\mu$m, $d=50\mu$m), and the averaging was performed over 105 realizations of the sample. The results of simulations indicate no magnetooptical effect and completely reciprocal localization.

\section{Transmission resonances}

Averaged localization decrement is associated with exponential decay of the incident wave deep into the infinite sample \cite{BaluniWillemsen,PingSheng,FreilikherGredeskul,BerryKlein}. For a finite sample, this is so only for \textit{typical} realizations. However, there exist some \textit{resonant} realizations of the sample at a given frequency (or, equivalently, resonant frequencies for a given realization) where transmission is  anomalously high and is accompanied by the accumulation of energy inside the sample. \cite{Frish,BliokhBliokhFreilikher} Such resonant transmission corresponds to excitation of the Anderson \textit{localized states} (quasi-modes) inside the sample. Akin to the resonant localized states in photonic crystal cavities, the transmission resonances in random structures are extremely sensitive to small perturbations: absorption, \cite{BliokhBliokhFreilikher} nonlinearity, \cite{Shadrivov} and, as we show here, magnetoactivity.

\begin{figure}[t]
\centering
\includegraphics[width=0.95\columnwidth]{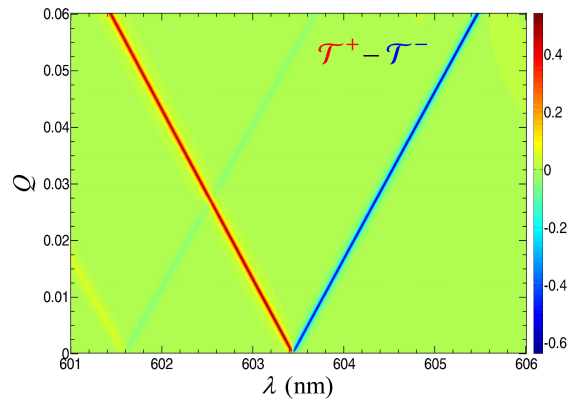}
\caption{\label{fig:faraday_diff_transm}
(Color online.) Differential transmittance, ${\mathcal T}^{+}-{\mathcal T}^{-}$, for two resonances from Fig.~4 as dependent on the value of magneto-optical parameter $Q$, cf. Eq.~(\ref{frequency shift-1}).}
\end{figure}

Figure~\ref{fig:faraday_spectrum} shows transmission spectra for two modes $\sigma=\pm 1$ (i.e., either with opposite helicities or propagation directions) in one realization of a magnetooptical sample in the Faraday geometry. The parameters of the sample are the same as in Section IIIA with $Q=0.06$. One can see strong splitting of the $\sigma=\pm 1$ transmission resonances which have exponentially narrow widths $\propto\kappa\exp (-\kappa N)/2\bar{w}$ \cite{BliokhBliokhFreilikher}. This offers strongly nonreciprocal, practically \textit{unidirectional}, propagation or polarization selectivity in the vicinity of resonant frequencies.

To estimate the splitting of resonances, we note that the wavenumbers in magnetooptical materials are shifted due to the Faraday effect, Eq.~(\ref{wavenumbers}). Hence, the shifts of the resonant wavenumbers of the random Faraday medium can be estimated by averaging of this shift over different materials in the structure:
%
\begin{equation}
\label{frequency shift-1}
\Delta k_{\rm res} \simeq \sigma\,\overline{\frac{qnk_0}{2} },
\end{equation}
where $\overline{(...)}$ stands for some average of $(...)$. Using $\overline{qn}\sim (q_a n_a + q_b n_b)/2$ for estimation in the two-component structure, we obtain $\Delta \lambda_{\rm res} \sim -\sigma \,3.6$ nm, which agrees with the $\sigma$-dependent splitting observed in Fig.~\ref{fig:faraday_spectrum}.

Figure~\ref{fig:faraday_diff_transm} displays the differential transmission for the waves with $\sigma=+1$ and $\sigma=-1$ as a function of magneto-optical parameter $Q$ for two resonances lying in a narrow frequency range in Fig.~\ref{fig:faraday_spectrum}. In agreement with estimation (\ref{frequency shift-1}), one observes the linear dependence of the resonance splitting on magnetization.

\begin{figure}[t]
\centering
\includegraphics[width=0.95 \columnwidth]{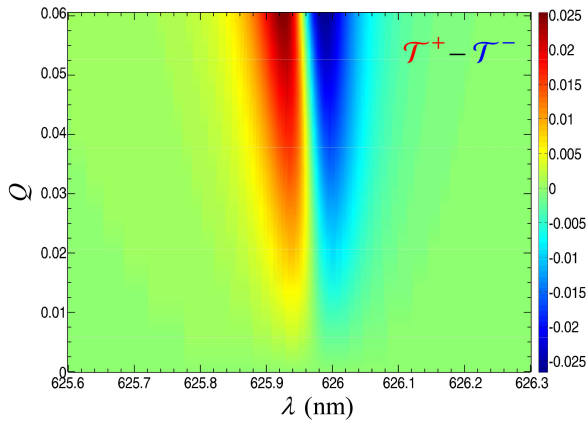}
\caption{\label{fig:Voigt_splitting}
(Color online.) Differential transmittance, ${\mathcal T}^{+}-{\mathcal T}^{-}$, for in the vicinity of a single resonance in the Voigt geometry (see Section IIIB for details) as dependent on the magneto-optical parameter $Q$.}
\end{figure}

In the Voigt geometry, the resonances also allow nonreciprocal transmission and demonstrate splitting of the resonant frequencies.
In Fig.~\ref{fig:Voigt_splitting} we show differential transmission for reciprocal waves with $\sigma=\pm 1$ in the vicinity of one resonance for the three-component structure considered in Section IIIB. The splitting is very small in this case, and $\sigma=+1$ and $\sigma=-1$ resonances overlap significantly. Because of this, the differential transmittance in Fig.~6 is tiny, its amplitude linearly grows with $Q$, whereas the frequency positions of its maximum and minimum correspond to the width of the original resonance and are practically unchanged.

Unlike the wave-number shift in the Faraday geometry, the noreciprocal shift of resonant frequencies in the Voight geometry arises from the phases of the amplitudes $A$, Eq.~(\ref{wavenumbers-1}). These phases are responsible for the phases of transmission coefficients between the layers and can be estimated as $\phi \sim q (\sigma k_{\perp})/(\upsilon k_{\parallel}) \equiv q \tan\theta$, where $\theta$ is the angle of propagation with respect to the $z$-axis. The phases accumulated at a layer effectively shift the wave numbers as $\upsilon \Delta k_{\parallel} =\Delta k \cos\theta\sim \phi/w$, where $w$ is the thickness of the layer. Averaging over different materials in the random layered structure, we estimate the nonreciprocal shift of the resonant wave number:
%
\begin{equation}
\label{frequency shift-2}
\Delta k_{\rm res} \sim \overline{\frac{q\sin\theta}{w\cos^2\theta}}=\sigma\,\overline{\frac{q|\sin\theta|}{w\cos^2\theta}}.
\end{equation}
This shift is $\sigma$-dependent, i.e., nonreciprocal, and much smaller than the Faraday-geometry shift (\ref{frequency shift-1}) as $k\bar{w}>kd\gg 2\pi$ in the short-wavelength limit, Eq.~(\ref{short-wave}). For the parameters in use, with $Q=0.06$, we have $\Delta \lambda_{\rm res} \sim -\sigma \,3\cdot 10^{-4}$nm, which agrees with the data plotted in Fig.~6.

\section{Conclusions}
\label{sec:Conclusion}

We have studied the transmission and localization of light in magnetoactive layered structures. An interplay between the Anderson localization and magnetooptical effects brings about various nonraciprocal phenomena in the transmission characteristics. We have analyzed the effects of the medium magnetization on the wave transmission in both Faraday and Voigt geometries in the short-wavelength limit.

Specifically, in the Faraday geometry the averaged localization length acquires the first-order magnetooptical corrections of the opposite signs for the opposite propagation directions or opposite circular polarizations of light. This leads to a broadband nonreciprocity or polarization selectivity in the typical exponentially small transmission observed in such structures in the regime of the Anderson localization. At the same time, random transmission resonances acquire significant nonreciprocal frequency shifts which result in efficient unidirectional propagation at the given resonant frequency. In the Voigt geometry, for the TM-polarized waves, the localization length is always reciprocal in the first-order approximation, whereas the transmission resonances show nonreciprocal frequency shifts but much smaller than those in the Faraday geometry.

Thus, we have observed that disorder-induced localization of light in random layered structures with magnetooptical materials demonstrates nonreciprocal features in both the averaged localization length and individual transmission resonances. Our results demonstrate that the Anderson localization can significantly enhance the magnetooptical effects, and this property can be employed for a design of novel types of efficient nonreciprocal devices which do no require periodicity and specially designed cavities.

\section*{ACKNOWLEDGEMENTS}
\label{sec:Acknowledgements}
We acknowledge fruitful discussions with A. B. Khanikaev. This work was supported by the European Commission
(Marie Curie Action), Science Foundation Ireland (Grant No. 07/IN.1/I906), and the Australian Research Council.


\end{document}